\documentclass{article}
\usepackage{graphicx} 

\title{Enhancement Pattern Mapping for Detection of Hepatocellular Carcinoma in Patients with Cirrhosis}

\begin{document}

\maketitle

\author{}

\section{Authors}{
Newsha Nikzad, Resident Physician, Department of Internal Medicine1,2,3*
David Thomas Fuentes, Associate Professor, Department of Imaging Physics2*
Millicent Roach, Research Assistant, Cancer Physics and Engineering Lab2
Tasadduk Chowdhury, Research Assistant, Cancer Physics and Engineering Lab2
Matthew Cagley, Research Assistant, Cancer Physics and Engineering Lab2
Mohamed Badawy, Research Assistant, Department of Imaging Physics2
Ahmed Elkhesen Resident Physician, Department of Internal Medicine4
Manal Hassan, Associate Professor, Department of Epidemiology2
Khaled Elsayes, Professor, Department of Abdominal Imaging2
Laura Beretta, Professor, Department of Molecular and Cellular Oncology2
Eugene Jon Koay, Associate Professor, Department of Radiation Oncology2
Prasun Kumar Jalal, Stan and Sue Partee Endowed Chair - Hepatology, Assistant Professor - Medicine and Surgery1 
1Baylor College of Medicine, Houston, TX
2The University of Texas MD Anderson Cancer Center, Houston, TX
3The University of Chicago Medical Center, Chicago, IL
4Texas Tech University Health Sciences Center, Lubbock, TX
*Authors share first co-authorship in order above.}

\section{Abstract}{Background and Aims: Limited methods exist to accurately characterize risk of malignant progression of liver lesions in patients undergoing surveillance for hepatocellular carcinoma (HCC). Enhancement pattern mapping (EPM) measures voxel-based root mean square deviation (RMSD) and improves the contrast-to-noise ratio (CNR) of liver lesions on standard of care imaging. This study investigates the utilization of EPM to differentiate between HCC versus benign cirrhotic tissue.
Methods: Patients with liver cirrhosis undergoing MRI surveillance at a single, tertiary-care hospital were studied prospectively. Controls (n=99) were patients without lesions during surveillance or progression to HCC. Cases (n=48) were defined as patients with LI-RADS 3 and 4 lesions who developed HCC within the study period. RMSD measured with EPM was compared to the signal from MRI arterial and portovenous (PV) phases. EPM signals of liver parenchyma between cases and controls were quantitatively validated on an independent patient set using cross validation. 
Results: With EPM, RMSD of 0.37 was identified as a quantitative cutoff for distinguishing lesions that progress to HCC from background parenchyma on pre-diagnostic scans with an area under the curve (AUC) of 0.83 (CI: 0.73-0.94) and a sensitivity, specificity, and accuracy of 0.65, 0.97, and 0.89, respectively. At the time of diagnostic scans, a sensitivity, specificity, and accuracy of 0.79, 0.93, and 0.88 was achieved with an AUC of 0.89 (CI: 0.82-0.96). EPM RMSD signals of background parenchyma in cases and controls were similar (case EPM: 0.22 ± 0.08, control EPM: 0.22 ± 0.09, p=0.8). 
Conclusions: EPM differentiates between HCC and non-cancerous parenchyma in a surveillance population and may aid in early detection of HCC. Future directions involve applying EPM for risk stratification of indeterminate lesions.
}

\section{Introduction}{The rapidly rising incidence of hepatocellular carcinoma (HCC) can be attributed to several risk factors, including metabolic syndrome, alcohol, viral hepatitis related to HBV or HCV, and other genetic and environmental etiologies. Morbidity and mortality related to HCC remain major challenges for the healthcare system throughout the world. Guidelines recommend surveillance for patients with cirrhosis for early detection of HCC to improve clinical outcomes. In patients with cirrhosis, the American Association for the Study of Liver Diseases (AASLD) recommends surveillance with abdominal ultrasound (US) every 6 months, with or without serum alpha-fetoprotein (AFP).1 Similarly, the European Association for the Study of the Liver (EASL) recommends surveillance with abdominal US every 6 months and notes the suboptimal cost-effectiveness of biomarkers such as AFP.2 US is an affordable, safe, and accessible imaging method; however, Tzartzeva et al. found that the sensitivity of US alone or with AFP for early-stage HCC is only 47
Suboptimal performance of US for small lesions motivates the use of contrast-enhanced magnetic resonance imaging (MRI),8 which detects smaller malignant lesions with more sensitivity than US (84
There is a need for new minimally invasive tools that better risk stratify patients under surveillance and detect HCC at earlier stages with higher sensitivity and specificity, given the limitations of US and MRI.1 While most current surveillance methods utilize clinical, demographic, and blood-based biomarkers, diagnostic methods utilize imaging, specifically computed tomography (CT) and MRI.10,13 The Liver Imaging Reporting and Data System (LI-RADS) represents an attempt to classify liver nodules for probability of malignancy with CT or MRI using a standardized method to minimize to discrepancies among radiologists.14 LI-RADs uses tumor size and arterial phase hyperenhancement as defining features for risk stratification, while major features of an enhancing capsule, washout, and threshold growth can further increase the confidence in a malignant or benign diagnosis. However, there is significant heterogeneity within the LI-RADS groups, especially for LR-3 and LR-4 lesions.1,14 Limited diagnostic performance is especially evident in these categories as 38
Recent studies have investigated machine learning and radiomic approaches for HCC detection, such as enhancement pattern mapping (EPM).10,11 EPM is a novel voxel-based signal analysis technique that quantifies the difference in enhancement over time of a given voxel in the liver compared to either a patient-specific or population-based normal liver model, providing a measurement and visualization of how different the signal is over an entire volume of interest. EPM expands on the available set of imaging features, and it provides an interpretable value that is based on angiogenesis and tumor perfusion that are fundamental to HCC pathophysiology improving its diagnostic performance. Previous medical literature has indicated that EPM algorithm improves the contrast-to-noise (CNR) ratio enhancement for lesion detection in hepatobiliary malignancy.12 Therefore, we embarked on this study to test the hypothesis that EPM can differentiate between HCC and cirrhotic parenchyma on pre-diagnostic and diagnostic MRI scans, with a future view of applications of EPM for early detection of HCC in patients undergoing surveillance. }

\section{Materials and Methods}{Patient selection and characteristics. 
With approval from the Baylor College of Medicine Institutional Review Board (H-47711 and H-45208) and MD Anderson Cancer Center (PA14-0646), all consecutive patients presenting with cirrhosis at the Hepatology Clinic between 2012 and 2020 at a single tertiary care hospital (Baylor St. Luke’s Medical Center) were prospectively followed. The institutional practice for HCC surveillance is with contrast-enhanced MRI and surveillance is performed every six months. Patients with cirrhosis were included in the study if they had at least two consecutive contrast-enhanced MRIs, utilizing liver protocol, for HCC surveillance. Patients with cirrhosis presenting with HCC in the initial scan were excluded.
Cases were defined as patients with a LI-RADS 3 or 4 lesion identified in a pre-diagnostic scan that progressed to HCC in the subsequent diagnostic scan. Controls did not have a lesion (LI-RADS 3 or more) identified on a single timepoint scan. Controls were age and sex-matched with cases to minimize confounding variables in patient characteristics. Patients were excluded after initial review if follow up imaging was unavailable (1/166 patients, 0.6

Figure 1. Patient selection process.

Statistical analysis to describe patient demographics was performed using SPSS Statistics (Version 26, SPSS Ltd, Chicago, IL). Data were tested for normality and homogeneity of variance using a Shapiro-Wilk test. Based on this outcome (P<0.05) and after visual examination of each variable’s histogram and QQ plot, data were reported as mean (SD) for normally distributed variables and as median (IQR 25th to 75th percentiles) for asymmetrical distributed data. A Student’s t-test was used to compare unpaired symmetrical continuous variables, and the non-parametric Mann-Whitney test was used for unpaired asymmetrical continuous variables. The Chi-square test or Fisher’s exact test was used to compare binary variables. A P<0.05 and a confidence interval (CI) of 95
Region of interest (ROI) Placement. 
Previous studies have demonstrated decreasing segmentation accuracy as a function of lesion size.17 Thus, one regions of interest was manually placed on each observed HCC lesion to avoid potential confounding factors from lesion segmentation inaccuracies. Similarly, to avoid any potential confounding factors from auto segmentation inaccuracies ROIs in background liver parenchyma were manually and randomly sampled. The normal liver ROIs were selected by visually analyzing the parenchyma of the liver to avoid medium to large blood vessels, cysts, and bile ducts. The software application ITK-SNAP12 was used to perform segmentation of ROIs on the arterial phase scan for all cases. A total of 8 ROIs were selected per image slice, with 3 slices per case, giving a total of 24 ROIs per case. The diameter of each ROI was selected as 6 millimeters. ROIs of lesions were delineated on the arterial phase of the contrast-enhanced scan. Arterial phase hyperenhancement of the lesion, venous and delayed phases washout, in addition to the lesion’s size and growth pattern were the criteria used to assign category codes based on LI-RADS version 2018 guidelines.14  
Enhancement pattern mapping. 
A three-dimensional, voxel-based method called the EPM algorithm was used for quantitative image analysis. Previous implementation of the EPM algorithm11 in multi-phase CT was modified for multi-phase MRI data. Briefly, the generalized enhancement pattern, such as the change in intensity values over the period of multi-phase MRI due to uptake and washout of contrast materials, of the liver was acquired from the registered multi-phase MRI scans. The normal liver enhancement curve was obtained by fitting the MRI intensity values over time within user-selected ROIs, as described in the previous section, sampled uniformly across normal liver parenchyma from the given patient. Second, the root-mean-square deviation (RMSD) for each voxel was computed by taking the average across all time points of the squares of the differences between the generalized normal liver intensity and the voxel intensity and then taking the square root of the average. Finally, the calculated RMSD values of all voxels were mapped to the original MRI coordinates.
The normal liver enhancement curve was obtained by fitting the MRI intensity values sampled from the normal liver ROIs over the period of contrast phases by a piece-wise smooth function, where each piece was a second-order polynomial. The EPM algorithm was implemented numerically using MATLAB (MathWorks, Inc.). Contrast to noise (CNR) measurements within the EPM image as well as the original multiphase MRI data were calculated as the average intensity value of the lesion minus the average intensity of healthy tissue divided by the standard deviation of the intensity of the health tissue. Here the intensity average and standard deviation were calculated over the ROI within the lesion and healthy tissue, as described in the previous section. A Wilcoxon rank-sum test is used to evaluate the statistical significance of the difference in the EPM RMSD measurements between cases and controls. RMSD differences between LI-RADS categories of the cases were also evaluated. 
ROC analysis was applied to study the EPM RMSD threshold to discriminate cases and control. The optimal cut point for the ROC analysis was defined as the point with the closest Euclidean distance to the perfect classifier (sensitivity=specificity=1). To validate the cut point, the variability of the optimal EPM cut point across 5-folds of the case and control dataset was evaluated. For each fold, the cut point was obtained from the training data independent from the validation hold-out fold. On average, within 5 fold cross validation, the data is split into 80

\section{Results}{Data curation. We identified consecutive patients with cirrhosis who underwent surveillance at our high-volume Hepatology clinic from 2012-2020. 58 cases developed HCC on surveillance and fulfilled the selection criteria, and 48 cases included in the study. 99 matched patients were designated controls.  Cases and controls were similar in baseline characteristics, as outlined in Table 1. The median age was 60 years (IQR 55-64) for combined cohort, 59 years (IQR 55-64) for controls, and 60 years (IQR 55-64) for cases. Most patients from both cohorts were male (n=89, 60.5

Table 1. Baseline characteristics of cases and controls.

EPM Analysis. On pre-diagnostic scans for cases, the mean CNR was as follows: 3.62 on EPM, 2.39 on arterial phase, and 1.05 on PV phase. Similarly, on diagnostic scans, the mean CNR was as follows: 3.58 on EPM, 2.35 on arterial phase, and 0.89 on PV phase. An example of the ROI including a lesion that was used to calculate the CNR of the image intensity on the arterial phase image and the EPM image is shown in Figure 2.

Figure 2. Representative ROI used for CNR analysis of the image intensity on EPM (left) and on arterial phase MRI (right) with lesion indicated. 

Boxplots of the EPM RMSD within lesions and background liver parenchyma is shown in Figure 3. Lesions in cases demonstrated a greater median EPM RMSD on pre-diagnostic scans (LI-RADS 3 and 4) and diagnostic scans (HCC) compared to parenchyma (p < 0.05). The average EPM RMSD of the background liver parenchyma in pre-diagnostic and diagnostic scans of the case patients and the single timepoint scan of control patients were not statistically different (control parenchymal ROI = 0.22 ± 0.09, case parenchymal ROI = 0.22 ± 0.08, p = 0.8).
 	    		
Figure 3. Box plots of EPM RMSD between observed lesions for pre-diagnostic and diagnostic scans in cases. (Left) On pre-diagnostic scans in cases, the median EPM RMSD observed was 0.44 for lesions and 0.22 for parenchyma. (Right) On diagnostic scans, the median EPM RMSD observed was 0.50 for lesions and 0.22 for parenchyma. (Left and right)

ROC analysis for EPM RMSD in differentiating cases and controls on pre-diagnostic and single timepoint scans is shown in Figure 4(a). A sensitivity, specificity, and accuracy of 0.65, 0.97, and 0.89 is achieved at the optimal threshold of EPM RMSD 0.37. The area under the curve (AUC) of EPM RMSD at the pre-diagnostic time point was 0.83 (CI: 0.73-0.94). In a multivariable logistic regression model, adjusting for BMI, age, sex, and diabetes status, the association between EPM and eventual HCC status, as all pre-diagnostic timepoints progressed to HCC, is significant (OR=8.08e3; 95
ROC analysis for EPM RMSD in differentiating cases and controls on diagnostic and single timepoint scans is shown in Figure 4(b). A sensitivity, specificity, and accuracy of 0.79, 0.93, and 0.88 was achieved at the optimal threshold of EPM RMSD 0.35. The AUC of EPM RMSD at the diagnostic time point was 0.89 (CI: 0.82-0.96). In a multivariable logistic regression model, adjusting for BMI, age, sex, and diabetes status, the association between EPM and HCC status is significant (OR=2.91e5; 95
Further permutations in ROC analysis for EPM RMSD in differentiating cases and controls on single timepoint and pre-diagnostic scans is shown in Figure 4(c). Five-fold cross validation was performed to estimate the out of sample performance on an independent test set. The range of optimum EPM RMSD threshold for discriminating case and control included 0.43, 0.32, .030, 0.39, and 0.37. Five-fold analysis achieved an aggregate sensitivity, specificity, and accuracy of 0.75, 0.92, and 0.88; respectively. ROC analysis for EPM RMSD in differentiating cases and controls on single timepoint and diagnostic scans is shown in Figure 4(d). Five-fold cross validation was performed to estimate the out of sample performance. The range of optimum EPM RMSD threshold for discriminating case and control included 0.45, 0.32, 0.35, 0.33, and 0.36. Five-fold analysis achieved an aggregate sensitivity, specificity, and accuracy of 0.83, 0.94, and 0.90; respectively.

Figure 4. ROC analysis of optimal threshold for discriminating case and control for pre-diagnostic and diagnostic time points. (a) In-sample ROC analysis for EPM RMSD in differentiating cases and controls on pre-diagnostic and single timepoint scans is shown. Similarly, (b) in-sample ROC analysis for EPM RMSD in differentiating cases and controls on diagnostic and single timepoint scans is shown. (c) Five-fold cross validation ROC analysis for EPM RMSD in differentiating cases and controls on pre-diagnostic and single timepoint scans is shown. Similarly, (d) Five-fold cross validation ROC analysis for EPM RMSD in differentiating cases and controls on diagnostic and single timepoint scans is shown.
}

\section{Conclusion}{This study investigated the use of a novel EPM technique for distinguishing HCC from background cirrhotic parenchyma, including non-malignant lesions. Our results suggested that EPM successfully differentiates between cancerous lesions and non-cancerous parenchyma in a surveillance population approximately six months before they were diagnosed with standard MRI protocol. EPM results show a significant increase in the CNR compared to the arterial and PV phase imaging on both pre-diagnostic and diagnostic scans. The CNR improvement on EPM is due exclusively to lesion signal amplification from calculated intensity differences between multiple phase scans. With EPM, a RMSD of 0.37 was a quantitative value to characterize lesions that progress to HCC on pre-diagnostic imaging (AUC of 0.83, CI: 0.73-0.94). The median time between pre-diagnostic and diagnostic scans for cases was 6.8 months (IQR 5.5-11.4). Further, the sensitivity (0.79), specificity (0.93), and accuracy (0.88) of EPM show an improvement during the diagnostics scan over the sensitivity (0.65), specificity (0.97), and accuracy (0.98) observed for the pre-diagnostic scan. 0.65, 0.97, and 0.89, respectively. The improvement in EPM performance agrees with intuition that lesions with higher LIRADs score would be expected to have a stronger EPM signal. 
Machine learning methods are a running theme to novel approaches for HCC detection and direct prediction of the LI-RAD score of a liver lesion. An overall accuracy of 60
Dissimilarly, this study investigated a neural network approach to differentiate lesions within the LR-3 and LR-4 categories, which are subject to greater heterogeneity and indeterminacy, from background cirrhotic parenchyma. The consistency in background signal between cases and controls suggests that the EPM RMSD value in liver parenchyma is unlikely to predict the future location of a new lesion. The current results indicate that an EPM RMSD cutoff of 0.37 would identify LR-3 and LR-4 lesions that progress to HCC. The range of cutoff values in five-fold cross validation agrees with this in-sample cutoff value. Sensitivity, specificity, and accuracy are comparable between in-sample and cross-validation analyses. The same analysis at the diagnostic time point provides a quantitative reference for the patients with known HCC. Parallels between pre-diagnostic and diagnostic values indicate that the EPM signal is likely to detect HCC earlier in the surveillance period. 
Dilation of the liver mask was important in achieving robust results. Generally, over-segmentation of the liver did not decrease the registration accuracy. However, under-segmentation of the liver was prone to more registration errors due to the registration algorithm not directly visualizing the entire liver for guidance. Adding the dilation effectively ensured that the liver was included in the mask used to guide the image registration. This approach further had the effect of reducing the sensitivity of the approach to segmentation errors. An additional limitation of this study is that the image analysis pipeline was not fully automated. Manual ROI were placed on lesions and liver parenchyma to facilitate EPM analysis. Our data did not include comparative analyses between different LI-RADS category lesions, as the primary goal of the study was to assess the foundational feasibility of using EPM to define malignant potential of indeterminant lesions. Additionally, we did not investigate lesions in LI-RADS M category, metastatic lesions, or cholangiocarcinoma. We also did not study the risk of progression to HCC based on cirrhosis etiology. 
In conclusion, EPM identifies quantifiable differences between HCC cases and controls in a population with cirrhosis under surveillance, and a threshold cutoff of 0.37 was found to be predictive approximately six months prior to diagnosis of HCC. Given the physiology integrated into this study’s EPM methodology, the findings highlight the potential applications of EPM as an imaging biomarker in the early detection realm. As lesion enhancement and transformation are intrinsically tied to visual evaluation by radiologists, this study provides a quantitative approach to the traditional qualitative approach of radiology. Future studies will include patients with LR-3 and LR-4 lesions without progression to HCC as a control group, and cohorts with a larger sample size will be diversified to have sex, race, ethnicity, and cirrhosis etiology distributions representative of the general population. This may subsequently allow for assessment of EPM in risk stratification for patients under surveillance for HCC, distinguishing which lesions are likely to transform into cancer. Our study introduces EPM as a prospective means of predicting lesion progression to malignancy, achieving early curative interventions, and individualizing care for patients at risk of development of HCC. }

\section{References}{1.	Marrero JA, Kulik LM, Sirlin CB, et al. Diagnosis, Staging, and Management of Hepatocellular Carcinoma: 2018 Practice Guidance by the American Association for the Study of Liver Diseases. Hepatology. Aug 2018;68(2):723-750. doi:10.1002/hep.29913
2.	European Association for the Study of the Liver. EASL Clinical Practice Guidelines: Management of hepatocellular carcinoma. J Hepatol. Jul 2018;69(1):182-236. doi:10.1016/j.jhep.2018.03.019
3.	Singal AG, Lampertico P, Nahon P. Epidemiology and surveillance for hepatocellular carcinoma: New trends. J Hepatol. Feb 2020;72(2):250-261. doi:10.1016/j.jhep.2019.08.025
4.	Tzartzeva K, Obi J, Rich NE, et al. Surveillance Imaging and Alpha Fetoprotein for Early Detection of Hepatocellular Carcinoma in Patients With Cirrhosis: A Meta-analysis. Gastroenterology. May 2018;154(6):1706-1718 e1. doi:10.1053/j.gastro.2018.01.064
5.	Giannini EG, Cucchetti A, Erroi V, Garuti F, Odaldi F, Trevisani F. Surveillance for early diagnosis of hepatocellular carcinoma: how best to do it? World J Gastroenterol. Dec 21 2013;19(47):8808-21. doi:10.3748/wjg.v19.i47.8808
6.	Hassett M, Yopp AC, Singal AG. Surveillance for hepatocellular carcinoma: how can we do better? Am J Med Sci. Oct 2013;346(4):308-13. doi:10.1097/MAJ.0b013e31828318ff
7.	Singal AG, Nehra M, Adams-Huet B, et al. Detection of hepatocellular carcinoma at advanced stages among patients in the HALT-C trial: where did surveillance fail? Am J Gastroenterol. Mar 2013;108(3):425-32. doi:10.1038/ajg.2012.449
8.	Simmons O, Fetzer DT, Yokoo T, et al. Predictors of adequate ultrasound quality for hepatocellular carcinoma surveillance in patients with cirrhosis. Aliment Pharmacol Ther. Jan 2017;45(1):169-177. doi:10.1111/apt.13841
9.	Kim HL, An J, Park JA, Park SH, Lim YS, Lee EK. Magnetic Resonance Imaging Is Cost-Effective for Hepatocellular Carcinoma Surveillance in High-Risk Patients With Cirrhosis. Hepatology. Apr 2019;69(4):1599-1613. doi:10.1002/hep.30330
10.	Mahmud N, Hoteit MA, Goldberg DS. Risk Factors and Center-Level Variation in Hepatocellular Carcinoma Under-Staging for Liver Transplantation. Liver Transpl. Aug 2020;26(8):977-988. doi:10.1002/lt.25787
11.	Park PC, Choi GW, M MZ, et al. Enhancement pattern mapping technique for improving contrast-to-noise ratios and detectability of hepatobiliary tumors on multiphase computed tomography. Med Phys. Jan 2020;47(1):64-74. doi:10.1002/mp.13769
12.	Yushkevich PA, Piven J, Hazlett HC, et al. User-guided 3D active contour segmentation of anatomical structures: significantly improved efficiency and reliability. Neuroimage. Jul 1 2006;31(3):1116-28. doi:10.1016/j.neuroimage.2006.01.015
13.	Goldberg DS, Taddei TH, Serper M, et al. Identifying barriers to hepatocellular carcinoma surveillance in a national sample of patients with cirrhosis. Hepatology. Mar 2017;65(3):864-874. doi:10.1002/hep.28765
14.	Chernyak V, Fowler KJ, Kamaya A, et al. Liver Imaging Reporting and Data System (LI-RADS) Version 2018: Imaging of Hepatocellular Carcinoma in At-Risk Patients. Radiology. Dec 2018;289(3):816-830. doi:10.1148/radiol.2018181494
15.	Lee YT, Wang JJ, Zhu Y, Agopian VG, Tseng HR, Yang JD. Diagnostic Criteria and LI-RADS for Hepatocellular Carcinoma. Clin Liver Dis (Hoboken). Jun 2021;17(6):409-413. doi:10.1002/cld.1075
16.	Zhang L, Wang JN, Tang JM, et al. VEGF is essential for the growth and migration of human hepatocellular carcinoma cells. Mol Biol Rep. May 2012;39(5):5085-93. doi:10.1007/s11033-011-1304-2
17.	Morshid A, Elsayes KM, Khalaf AM, et al. A machine learning model to predict hepatocellular carcinoma response to transcatheter arterial chemoembolization. Radiol Artif Intell. Sep 2019;1(5)doi:10.1148/ryai.2019180021
18.	Yamashita R, Mittendorf A, Zhu Z, et al. Deep convolutional neural network applied to the liver imaging reporting and data system (LI-RADS) version 2014 category classification: a pilot study. Abdom Radiol (NY). Jan 2020;45(1):24-35. doi:10.1007/s00261-019-02306-7
19.	Mokrane FZ, Lu L, Vavasseur A, et al. Radiomics machine-learning signature for diagnosis of hepatocellular carcinoma in cirrhotic patients with indeterminate liver nodules. Eur Radiol. Jan 2020;30(1):558-570. doi:10.1007/s00330-019-06347-w
20.	Wu Y, White GM, Cornelius T, et al. Deep learning LI-RADS grading system based on contrast enhanced multiphase MRI for differentiation between LR-3 and LR-4/LR-5 liver tumors. Ann Transl Med. Jun 2020;8(11):701. doi:10.21037/atm.2019.12.151
21.	Yasaka K, Akai H, Abe O, Kiryu S. Deep Learning with Convolutional Neural Network for Differentiation of Liver Masses at Dynamic Contrast-enhanced CT: A Preliminary Study. Radiology. Mar 2018;286(3):887-896. doi:10.1148/radiol.2017170706
22.	Hamm CA, Wang CJ, Savic LJ, et al. Deep learning for liver tumor diagnosis part I: development of a convolutional neural network classifier for multi-phasic MRI. Eur Radiol. Jul 2019;29(7):3338-3347. doi:10.1007/s00330-019-06205-9

}

\end{document}